\documentclass[rnote]{aa} 

\usepackage{graphicx}
\usepackage{txfonts}
\begin{document}

   \title{Determination of the cross-field density structuring in coronal waveguides using the damping of transverse waves}

\titlerunning{Determination of transverse inhomogeneity in coronal waveguides}
\authorrunning{Arregui \& Asensio Ramos}

   \author{I. Arregui
          \inst{1,2}
          \and
          A. Asensio Ramos\inst{1,2}
          }

   \institute{Instituto de Astrof\'{\i}sica de Canarias, E-38205 La Laguna, Tenerife, Spain\\
              \email{iarregui@iac.es}
         \and
            Departamento de Astrof\'{i}sica, Universidad de La Laguna, E-38206 La Laguna, Tenerife, Spain\\
             }

   \date{Received ; accepted}

 
  \abstract
   {Time and spatial damping of transverse magnetohydrodynamic (MHD) kink oscillations is a source of information on the cross-field variation of the plasma density in coronal waveguides. }
   {We show that a probabilistic approach to the problem of determining the density structuring from the observed damping of transverse oscillations enables us to obtain information on the two parameters that characterise the cross-field density profile.}
   {The inference is performed by computing the marginal posterior distributions for density contrast and transverse inhomogeneity length-scale using Bayesian analysis and damping ratios for transverse oscillations under the assumption that damping is produced by resonant absorption.}
   {The obtained distributions show that, for damping times of a few oscillatory periods, low density contrasts and short inhomogeneity length scales are more plausible in explaining observations.}
   {This means that valuable information on the cross-field density profile can be obtained even if the inversion problem, with two unknowns and one observable, is a mathematically  ill-posed problem.}


   \maketitle
%

\section{Introduction}


Transverse magnetohydrodynamic (MHD) kink oscillations have been reported in numerous observations of solar coronal magnetic and plasma structures. First revealed in coronal loop observations using the Transition Region and Coronal Explorer (TRACE) by \cite{aschwanden99} and \cite{nakariakov99}, they also seem to  be an important part of the dynamics of chromospheric spicules and mottles \citep{depontieu07,kuridze13}; soft X-ray coronal jets \citep{cirtain07}; or prominence fine structures \citep{okamoto07,lin09}. 
Their presence over extended regions of the solar corona \citep{tomczyk07} may have implications on the role of waves in coronal heating.  Observations with instruments such as AIA/SDO, CoMP, and Hi-C by e.g., \cite{morton13} and \cite{threlfall13} have allowed us to analyse transverse oscillations with unprecedented detail.

The potential use of transverse oscillations as a diagnostic tool to infer otherwise difficult to measure coronal magnetic and plasma properties was first demonstrated by \cite{nakariakov01}, by interpreting them as MHD kink modes of magnetic flux tubes. Since then, a number of studies have used  seismology diagnostic tools that make use of oscillation properties, such as periods and damping times, to obtain information on the magnetic field and plasma density structuring \citep[see e.g., ][]{andries05b,arregui07a,verth08b,verth11}. An overview of recent seismology applications  can be found in \cite{arregui12b} and \cite{demoortel12}. 
The increase in the number of observed events has lately enabled the application of statistical techniques to seismology diagnostics  
\citep{verwichte13, asensioramos13}.

Some of these studies are concerned with the determination of the cross-field density structuring in waveguides supporting MHD oscillations. \cite{goossens02a} were the first to note that measurements of the damping rate of coronal loop oscillations together with the assumption of resonant absorption as the damping mechanism could be used to obtain estimates for the transverse inhomogeneity length scale.  By assuming a value for the ratio of the plasma density between the interior of the loop and the corona, they computed the inhomogeneity length scale for a set of 11 loop oscillation events. \cite{verwichte06} analysed how information on the density profile across arcade-shaped models can be obtained from the oscillation properties of vertically polarised transverse waves. When no assumption is made on the density contrast, \cite{arregui07a} and \cite{goossens08a} showed that an infinite number of
equally valid equilibrium models is able to reproduce observed damping rates, although they must follow a particular one-dimensional curve in the two-dimensional parameter space of unknowns. 
More recently, \cite{arregui11b} have shown how information on density contrast from observations can be used as prior information in order to fully constrain the transverse density structuring of coronal loops. \cite{arregui13b} have shown that the existence of two regimes in the damping time/spatial scales would enable the constraint of both the density contrast and its transverse inhomogeneity length scale. The feasibility of such a measurement has yet to be confirmed by observations.

We present the Bayesian solution to the problem, that makes use of the computation of marginal posteriors, and show that valuable information on the cross-field density profile can be obtained even if the inversion problem, with two unknowns and one observable, is a mathematically  ill-posed problem.


\section{Conditional probability and marginal posteriors}\label{procedure}

Consider the determination of a set of parameters {\boldmath$\theta$} related to a  theoretical model, $M$, that are compared to observed data, $d$.
Bayes rule for conditional probability \citep{bayes73} states that the probability of \mbox{\boldmath$\theta$} taking on given values, conditional on the observed data, the posterior probability $p(\mbox{\boldmath$\theta$} | d)$, is a combination of how well the data are reproduced by the model parameters, the likelihood function $p(d| \mbox{\boldmath$\theta$})$, and the probability of the parameters independently of the data, the prior distribution $p(\mbox{\boldmath$\theta$})$. These quantities are related as follows

\begin{equation}\label{bayes}
p(\mbox{\boldmath$\theta$} | d,M)=\frac{p(d | \mbox{\boldmath$\theta$},M)p(\mbox{\boldmath$\theta$},M)}{\int p(d|\mbox{\boldmath$\theta$}, M)p(\mbox{\boldmath$\theta$}, M)d\mbox{\boldmath$\theta$}},
\end{equation}

\noindent
and we have made explicit that all quantities are conditional on the assumed model $M$. The denominator is the so-called evidence,  an integral of the likelihood over the prior distribution that normalises the likelihood and turns it into a probability. The prior and likelihood represent probabilities that are directly assigned, whilst the posterior is computed. Probability in this context means the grade of belief on a statement about the value a parameter can take on, conditional on observed data. The resulting posterior is a distribution that quantifies this grade of belief.

Once the full posterior is known, the so-called marginal posterior enables us to calculate how a particular parameter of interest, $\theta_i$, is constrained by observations, by just performing an integral of the posterior over the rest of the model parameters

\begin{equation}\label{marginal}
p(\theta_i|d) = \int p(\theta | d) d\theta_1 \ldots 
d\theta_{i-1} d\theta_{i+1} \ldots d\theta_{N}.
\end{equation}

\noindent
This quantity encodes all information for model parameter $\theta_i$ available in the priors and the data and correctly propagates uncertainty in the rest of the parameters to the one of interest.

Consider an observable quantity, $c$, which, according to some theoretical model with parameters $a$ and $b$ is predicted to be
related to them by a given forward model, $c_{\rm model}=f(a,b)$, indicating that $c$ is a function of $a$ and $b$. A given value of $c$ can be obtained from different combinations of $a$ and $b$. The observational measurement of $c$ provides us with many equally valid combinations of $a$ and $b$ that give the observed $c$. In a real application, measurements are corrupted by noise, which makes necessary a probabilistic approach to the inversion problem. 

\begin{figure}
   \includegraphics[width=0.5\textwidth]{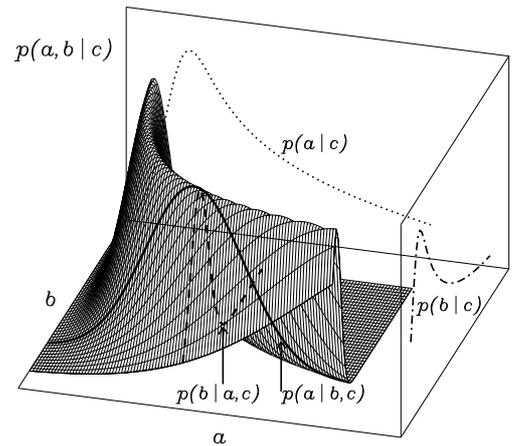}
   \caption{Surface plot of the joint probability $p(a,b | c)$ of parameters $a$ and $b$ for a given observation of $c$ according to the forward model $c_{\rm model}=a\cdot b$ and assuming a Gaussian likelihood (Eq.~[\ref{like1}]).  Conditional probabilities for selected values of parameters are indicated by arrows. Marginal posteriors are projected onto the two vertical planes.}\label{joint}
\end{figure}

In our analysis, we consider that observations are corrupted with Gaussian noise and that they are statistically independent. Then, the observed values of $c$ and the theoretical predictions can be compared by adopting a Gaussian likelihood of the form

\begin{equation}\label{like1}
p(c|a,b) = \frac{1}{\sqrt{2\pi}\sigma_c} \exp \left\{-\frac{\left[c-c_{\mathrm{model}}(a,b)\right]^2}{2 \sigma^2_c} \right\},
\end{equation}
with $\sigma_c$ the uncertainty associated to the measured $c$. We also assume uniform prior distributions for $a$ and $b$ over given ranges, that are irrelevant to our current discussion. The actual capability of the model to reproduce the observation can be evaluated by computing the joint probability of $a$ and $b$, conditional on $c$. Figure~\ref{joint} displays this two-dimensional distribution, in the case the forward model is simply the product of $a$ and $b$. This forward model is chosen here for simplicity. Any other choice  is equally valid to describe our approach. At each position, the magnitude of the joint probability $p(a,b | c)$ inform us on the ability of that particular combination of parameters to reproduce a particular value for the observed $c$.

A cut on this surface along the direction of the parameter $b$, at a particular value of the parameter $a$,  is the probability of $b$ conditional on $a$ and the observation $c$, $p(b | a, c)$. A cut for another value for $a$ will results in a different probability distribution for $b$.  The full probability of $b$, conditional only on data is the integral over all possible values for $a$, i.e., the marginal posterior $p(b | c)$.  Correspondingly, a cut along the direction of the parameter 
$a$, at a particular value of the parameter $b$,  is the probability of $a$, conditional on $b$ and $c$, $p(a | b, c)$. The full probability of $a$, conditional on data is the integral over all possible values for $b$, i.e., the marginal posterior $p(a | c)$. The two marginal posteriors for $a$ and $b$ are shown on the vertical planes in Figure~\ref{joint}. They indicate that even if the observed $c$ can be reproduced by many combinations (product of values) of $a$ and $b$, some parameter values are more plausible than others. The particular result will depend on the forward model, which determines the shape of the joint distribution,  the data value with its associated uncertainty, and the range over which parameters are allowed to vary (the priors). 


\section{The probability of a damping ratio}

As an application of the procedure outlined above, we consider the damping of transverse MHD kink waves by resonant absorption
in one-dimensional density tube models for coronal waveguides. The classic analysis in the thin tube and thin boundary approximations  \citep{goossens92a,ruderman02} leads to the following expression for the damping ratio

\begin{equation}\label{dratio}
r_{\rm model}=\frac{\tau_d}{P}=\frac{2}{\pi} \left(\frac{R}{l}\right) \left(\frac{\zeta+1}{\zeta-1}\right),
\end{equation}

\noindent
with $P$ and $\tau_{\rm d}$  the period and damping time of the oscillation, $\zeta=\rho_{\rm i}/\rho_{\rm e}$ the density contrast, and $l/R$ the transverse 
inhomogeneity length scale in units of the tube radius $R$. The factor $2/\pi$ arises from the assumed sinusoidal variation of the density profile at the tube boundary. The impact of alternative density profiles on seismology estimates is discussed in \cite{soler14b}. 
A similar expression is valid for propagating kink waves upon replacement of the damping time $\tau_{\rm d}$ by the damping length $L_{\rm d}$ and of the oscillation period $P$ by the longitudinal wavenumber $k_{\rm z}$, as shown by \cite{terradas10}.

In our particular application, the two unknown parameters {\boldmath$\theta$}=($\zeta$, $l/R$) will be inferred using the damping ratio as an observable, $d=r$, assuming the resonant damping model, $M$, as the explanation for the decay of the oscillations.  
We proceed as in Section~\ref{procedure}  and assign direct probabilities for the likelihood function and the prior distribution. We adopt a Gaussian likelihood function that relates the observed damping ratio, $r$, and the predictions of the model, $r_{\rm model}$, given by Equation~(\ref{dratio}), so that

\begin{equation}
p(r|\zeta, l/R) = \frac{1}{\sqrt{2\pi}\sigma}  \exp \left\{-\frac{\left[r-r_{\mathrm{model}}(\zeta, l/R)\right]^2}{2 \sigma^2} \right\},
\end{equation}

\noindent
with $\sigma$ the uncertainty associated to the measured damping ratio. We also adopt uniform prior distributions for both unknowns over given ranges, so that all the values inside those ranges are equally probable a priori, so we consider

\begin{equation}
p(\theta_i)=\frac{1}{\theta^{max}_i-\theta^{min}_i} \mbox{\hspace{0.4cm}} \mbox{for}  \mbox{\hspace{0.4cm}} \theta^{min}_i\leq\theta_i\leq\theta^{max}_i,
\end{equation}
and zero otherwise. Application of Bayes rule (Equation~[\ref{bayes}]) provides us with the full posterior, $p(\mbox{\boldmath $\theta$}|d)$, from which the marginal posteriors are obtained through marginalisation,

\begin{equation}
\begin{array}{ll}
p(\zeta | r)=\int p(\zeta,l/R | r) \ d(l/R), \\\\
 p(l/R | r)=\int p(\zeta,l/R | r) \ d\zeta.\\
 \end{array}
\end{equation}

\begin{figure}
   \centering
   \includegraphics[width=0.4\textwidth]{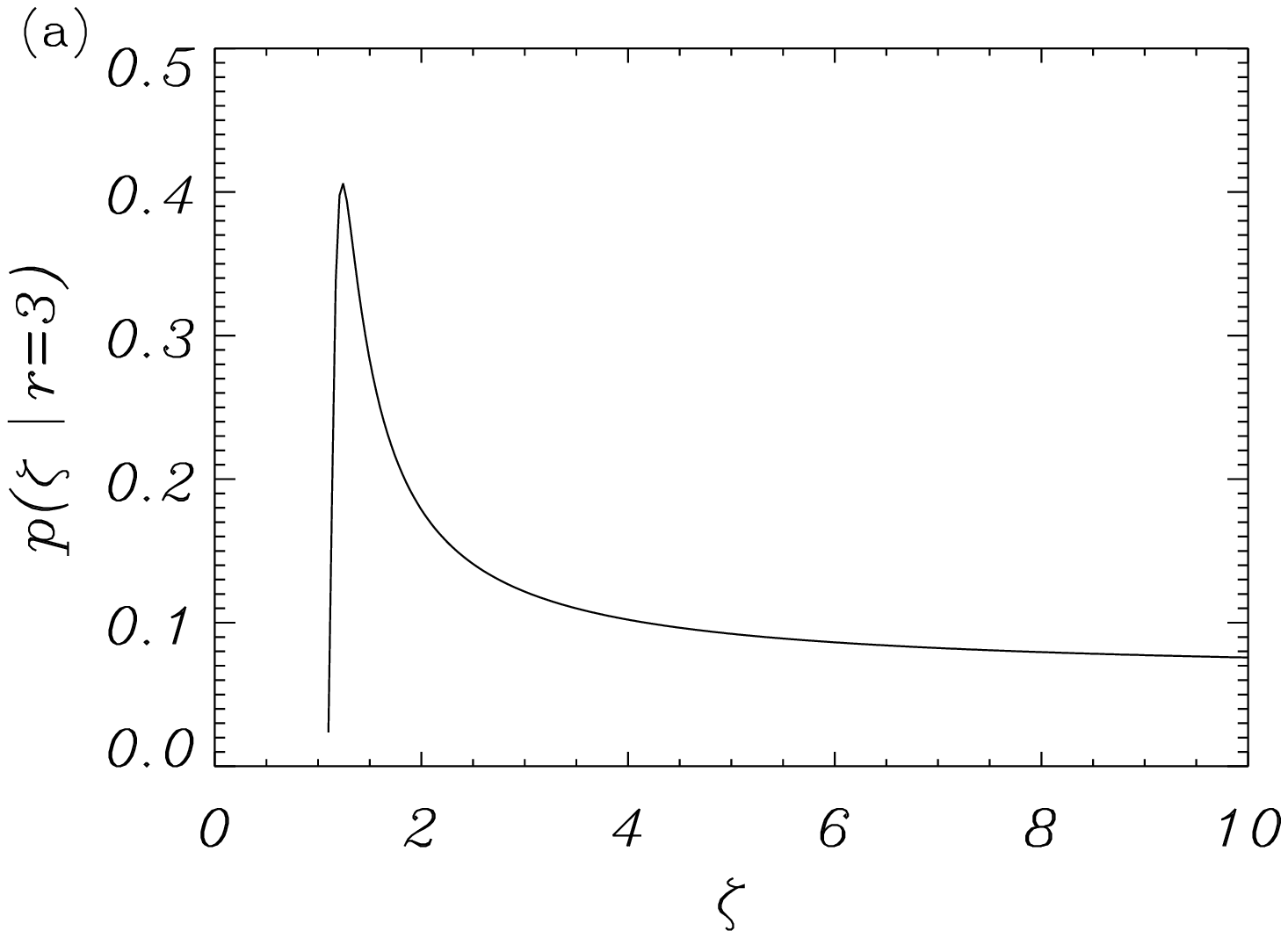}\\
   \includegraphics[width=0.4\textwidth]{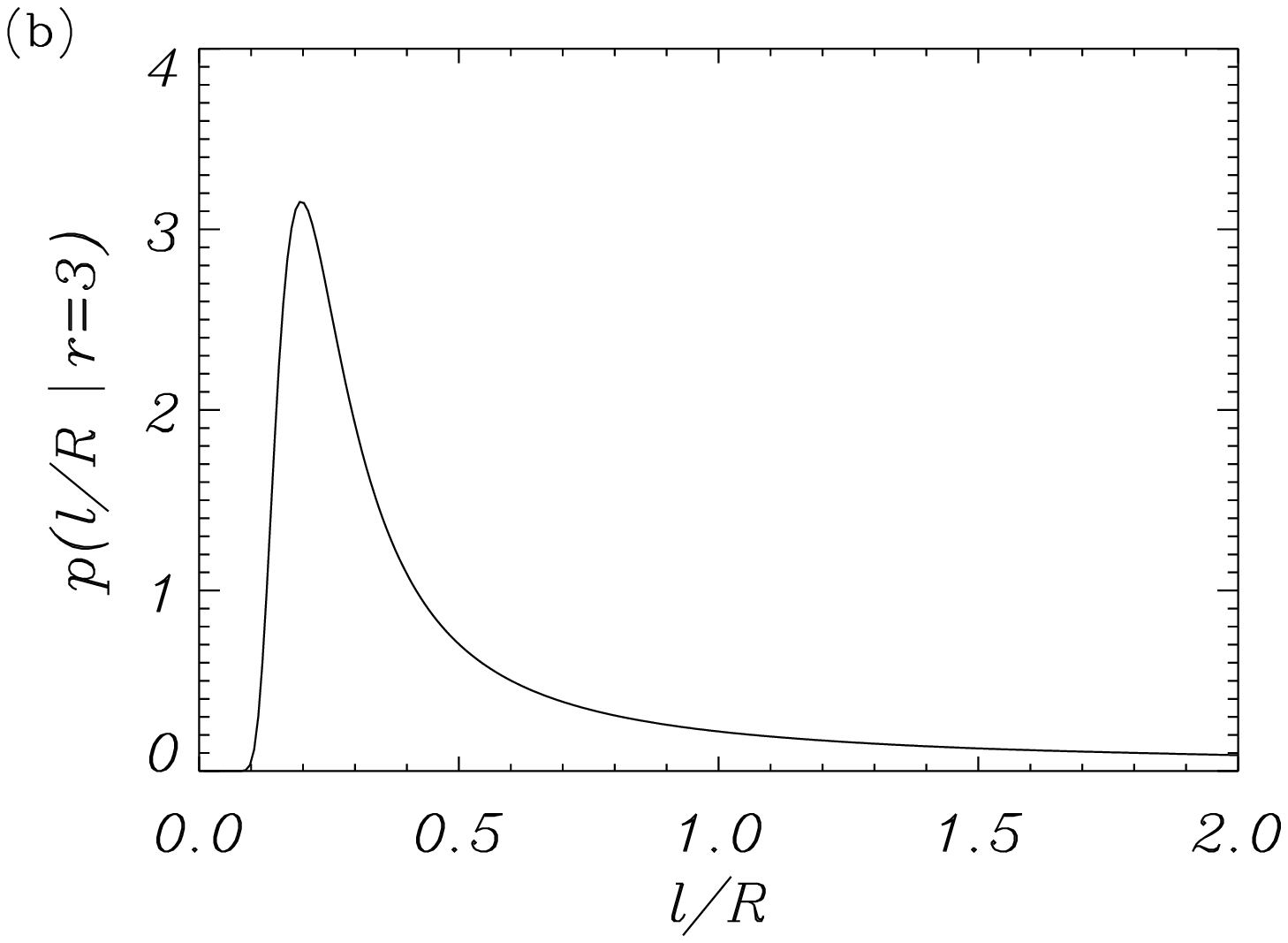}\\
   \includegraphics[width=0.4\textwidth]{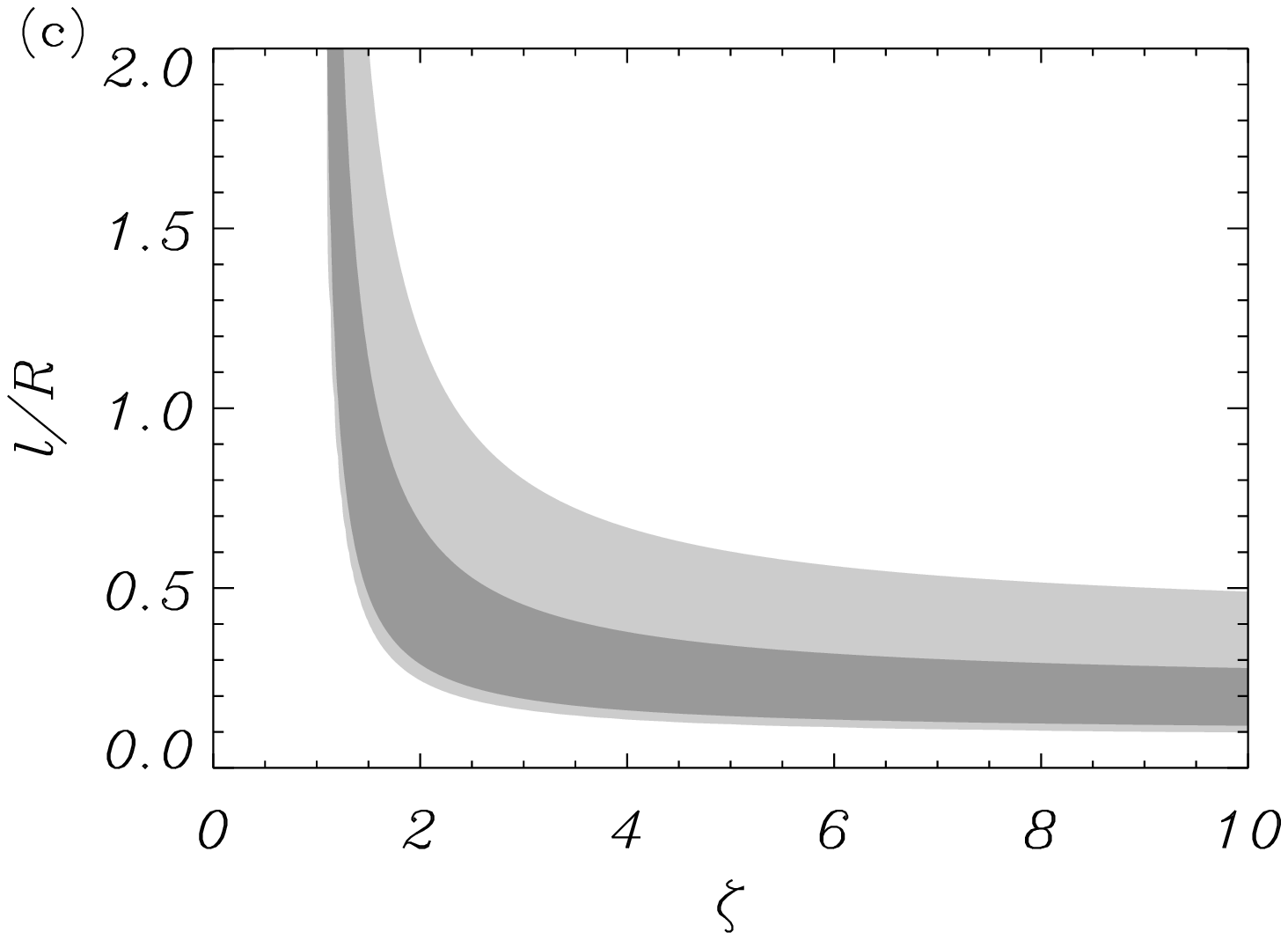}
   \caption{(a) and (b) Marginal posteriors for $\zeta$ and $l/R$ for an inversion with a measured damping ratio $r=3$ and $\sigma=1$, using uniform priors in the ranges $\zeta\in[1.1-10]$ and $l/R\in[0.01-2]$. (c) Joint two-dimensional posterior distribution for the two inferred parameters. The light and dark grey shaded regions indicate the 95\% and 68\% credible regions, respectively.}
              \label{inference}%
    \end{figure}

Figure~\ref{inference} shows an example inversion result. The inversion suggests that low density contrast values are preferred over large contrast ones. However, the posterior for density contrast displays a long tail, which means that this parameter can only be constrained with a large uncertainty.  A more constrained marginal posterior is obtained for the transverse inhomogeneity length scale, which points to short values of $l/R$ to be more plausible than models with a fully non-uniform layer. The joint two-dimensional distribution shows the combined posterior distribution for both parameters with the  68\% and 95\% credible regions.

\begin{figure*}
   \centering
   \includegraphics[width=0.4\textwidth]{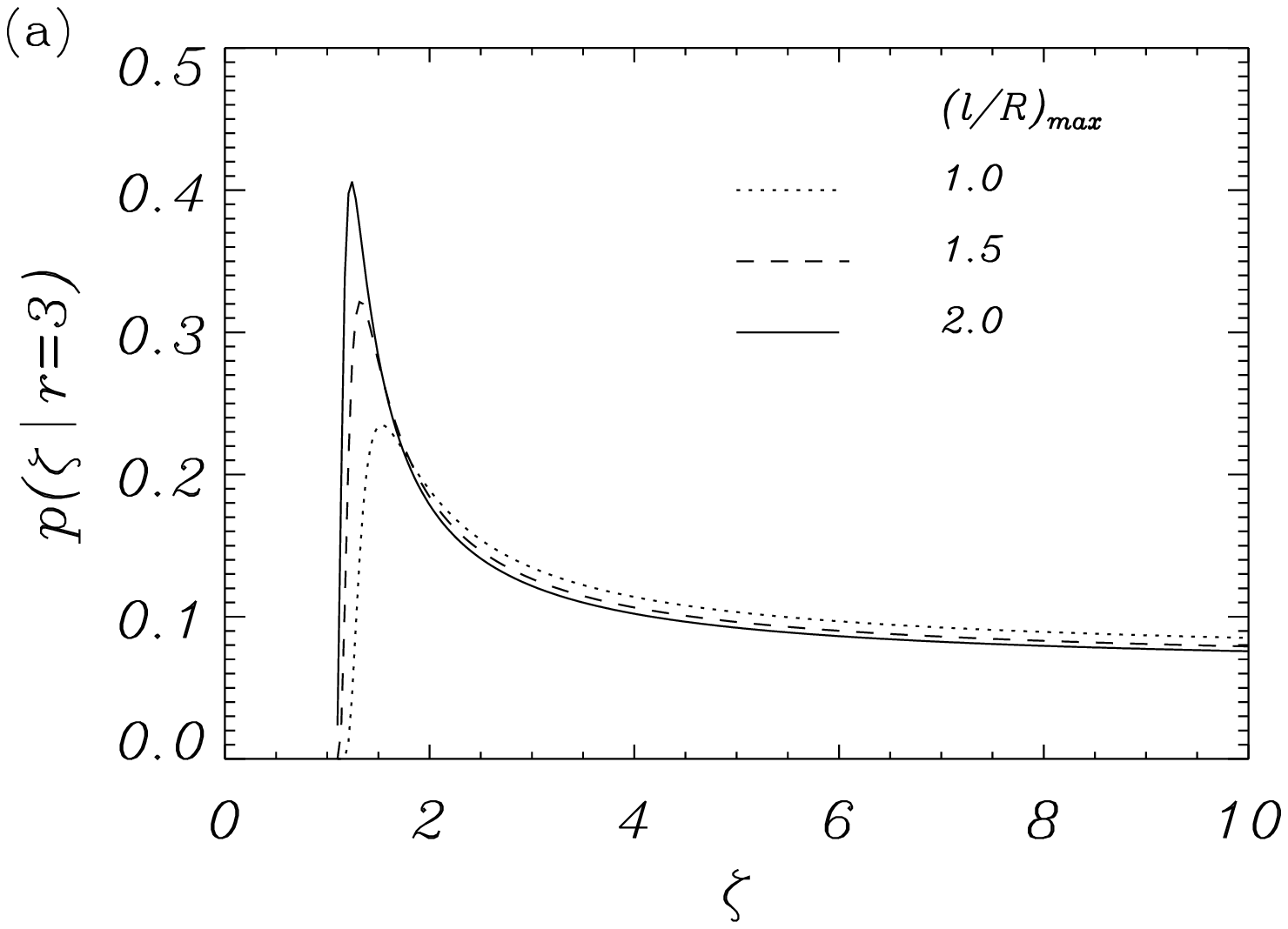}
   \includegraphics[width=0.4\textwidth]{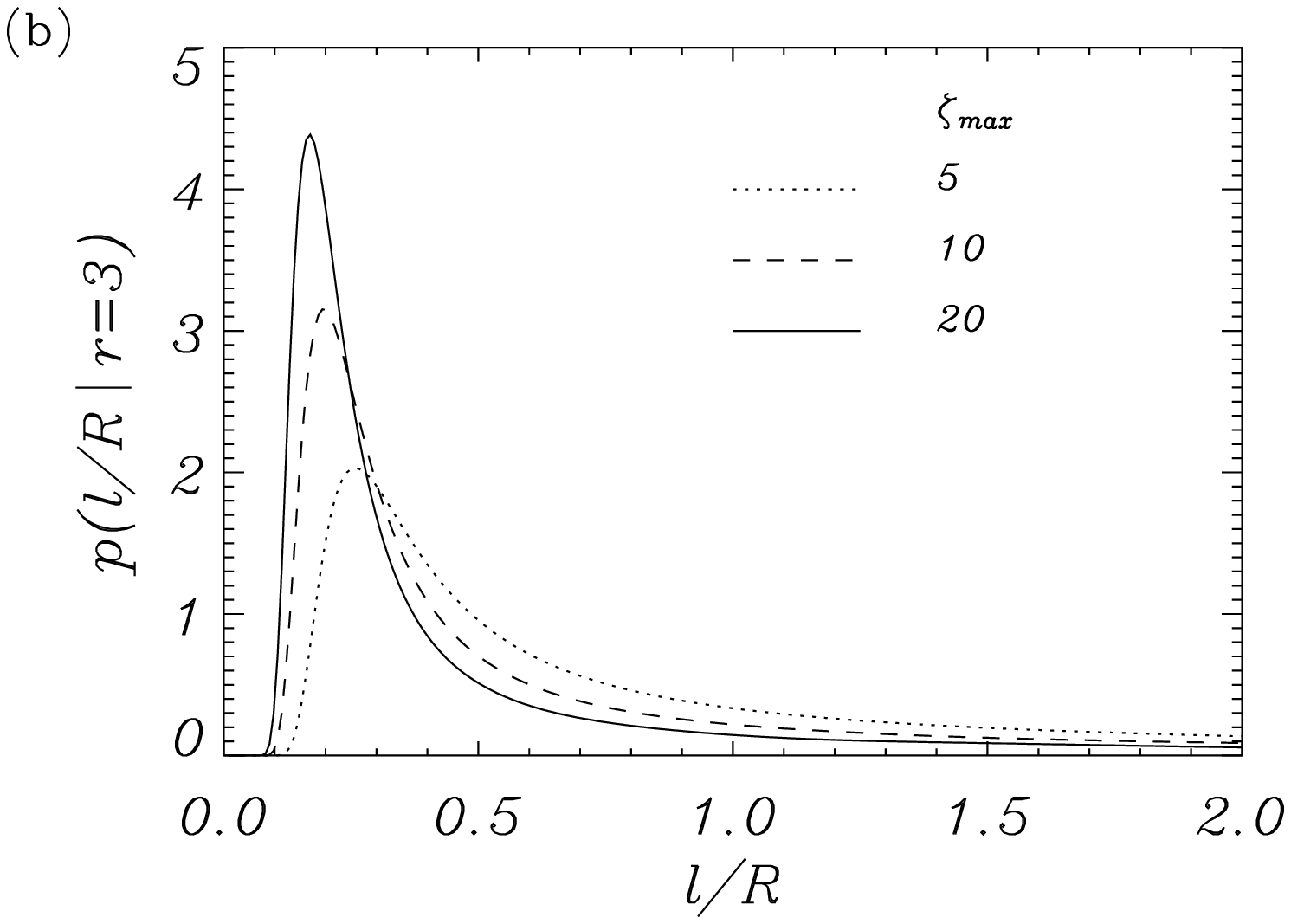}\\
   \includegraphics[width=0.4\textwidth]{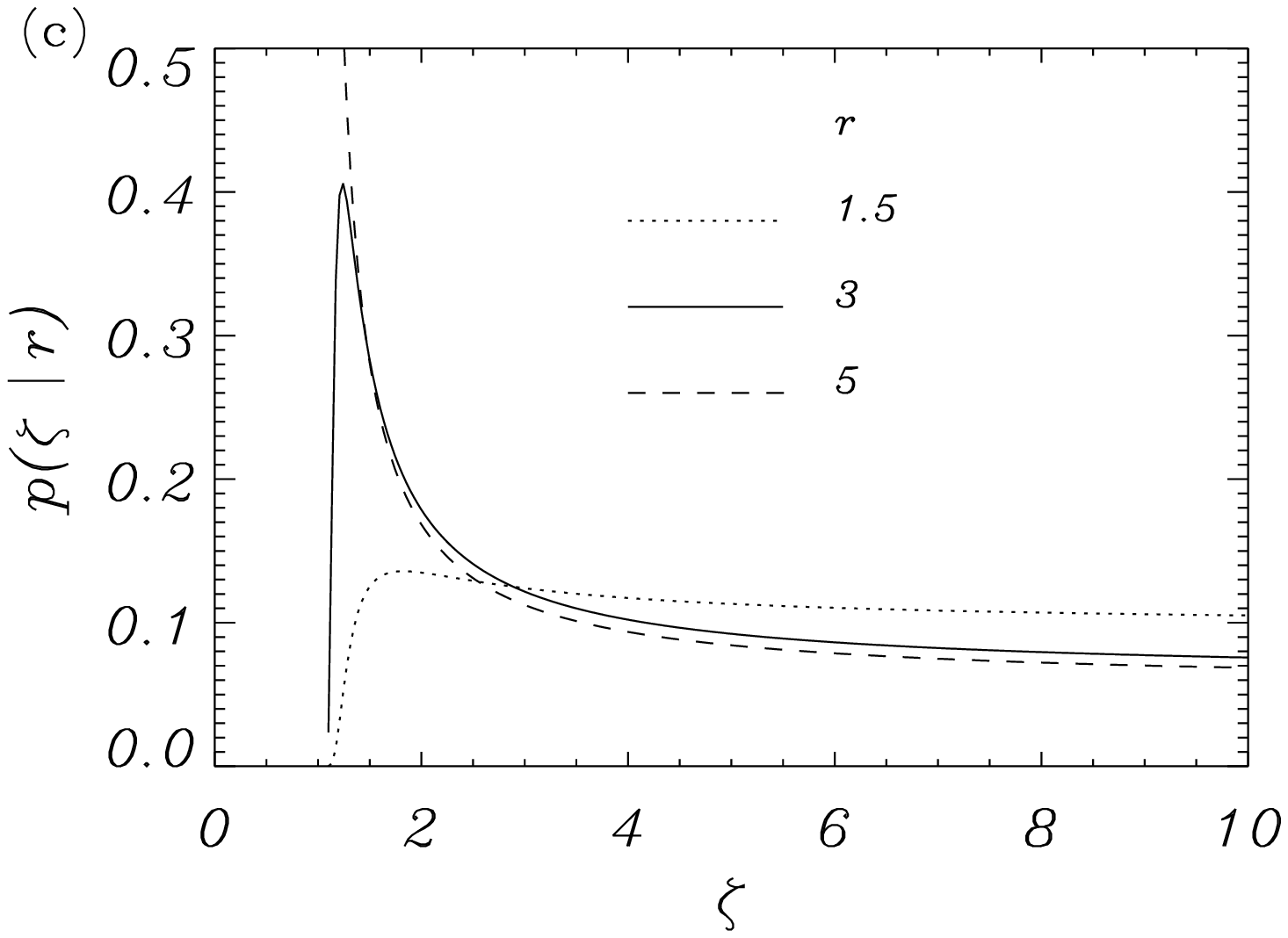}
      \includegraphics[width=0.4\textwidth]{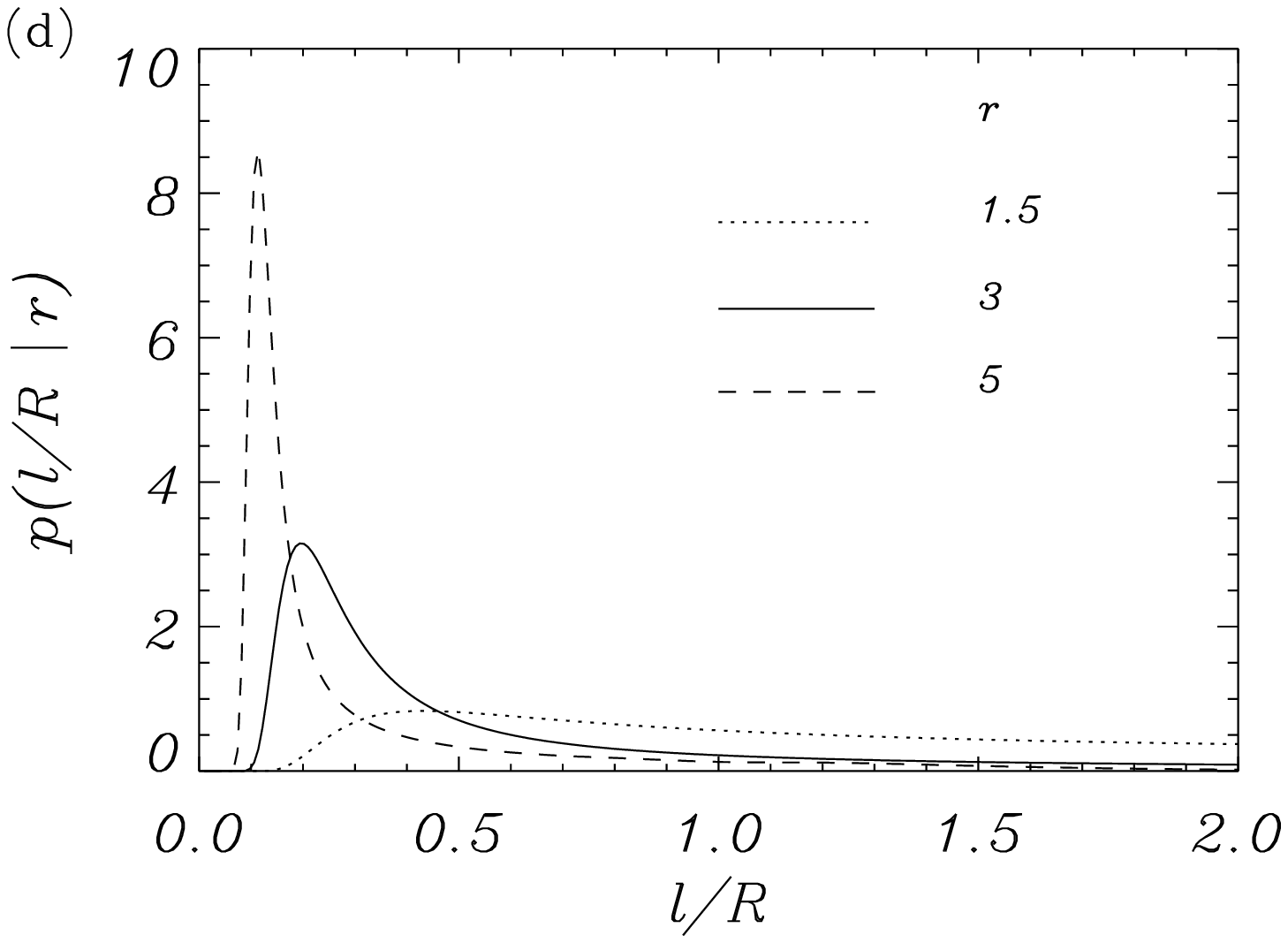}
   \caption{(a) and (b) Effect of different prior ranges with $l/R\in[0.01,(l/R)_{\rm max}]$ on the marginal posterior for $\zeta$ and with 
   $\zeta\in[1.1,\zeta_{\rm max}$] on the marginal posterior for $l/R$. (c) and (d)  Effect of three different damping ratio measurements on the inferred marginal posteriors.  In all figures, $\sigma=1$.}
              \label{diffs}%
    \end{figure*}

Inferences depend on the assumed prior ranges, since the integrals  run over the assumed parameter ranges. The values for $\zeta_{\rm min}$ and ($l/R$)$_{\rm min}$ were chosen so that they are slightly above the minimum values permitted by the theoretical model. Figure~\ref{diffs}a shows that varying the upper limit of the transverse inhomogeneity length scale in its prior distribution does not affect much the determination of density contrast. Alternatively, varying the upper limit of  the considered density contrast in its prior distribution results in marginal posteriors for $l/R$ that shift towards shorter transverse inhomogeneity length scales being more plausible (Figure~\ref{diffs}b).

Observed damping ratios in e.g., coronal loop oscillations are roughly in between 1 and 5. We have performed the inference for three particular values (see Figures~\ref{diffs}c and d).  We find that damping ratios slightly larger than one do not enable us to constrain the unknown parameters  using this method (dotted lines). Once the damping time is several times the period, e.g., $r=3$, well constrained distributions are obtained (solid lines). Finally, larger damping ratios still enable us to constrain $l/R$, but the only statement that can be made regarding the density contrast is that low values are more plausible than larger ones (dashed lines). 
In all our computations, with damping times of a few oscillatory periods, short transverse inhomogeneity length scales are always found as the most plausible ones.


\section{Conclusions}

Seismology of transverse MHD kink oscillations offers a way for obtaining information on the plasma density structuring
across magnetic waveguides. For resonantly damped kink mode oscillations, the determination of the cross-field density 
profile from observed damping ratios consists on the solution of an ill-posed mathematical problem with two unknowns and one observable

In this study we have introduced a modified Bayesian analysis technique that makes use of the basic definition of marginal posteriors, which are obtained not by sampling the posterior using a Markov Chain Montecarlo technique as in \cite{arregui11b}, but by performing the required integrals over the parameter space once the joint probability is computed. This has led to a better understanding about when and how the unknown parameters may be constrained.

The application of this technique for the computation of  the probability distribution of the unknowns enables us to draw conclusions 
about the most plausible values, conditional on observed  data. The procedure makes use of all available information and offers correctly
propagated uncertainty. Considering typical ranges for the possible values of the unknown parameters we find that, for damping times of a 
few oscillatory periods, low values for the density contrast are favoured and larger values of this parameter become less plausible. 
Regarding the transverse inhomogeneity length scale, short values below the scale of the tube radius are found to be more plausible than
larger length-scales near the limit for fully non-uniform tubes.

The procedure described here can be followed to obtain the most plausible values for the two unknowns that determine 
the cross-field density profile in coronal waveguides, upon assuming resonant absorption as the damping mechanism operating in transverse 
kink waves, and conditional on the observed damping ratios with their associated uncertainty.

In contrast to \cite{arregui11b}, who used two observables to determine three unknowns, we use one observable to determine two unknown parameters. The reason why we left out the Alfv\'en travel time is because of our interest on the cross-field density profile that is determined by $\zeta$ and $l/R$. The 
same technique presented in this study can be used by increasing by one the number of observables/unknowns.  We have found that this leads to 
similar posteriors for $\zeta$ and  $l/R$, with the additional information on the Alfv\'en travel time being also available.

\begin{acknowledgements}
 We acknowledge financial support by the Spanish Ministry of Economy and Competitiveness through projects AYA2010--18029 (Solar Magnetism and Astrophysical Spectropolarimetry), AYA2011--22846 (Dynamics and Seismology of Solar Coronal Structures), and the Consolider-Ingenio 2010 CSD2009-00038 project. We also acknowledge financial support through our Ram\'on y Cajal fellowships. We are grateful to the referee and to Ram\'on Oliver for comments and suggestions that improved the paper.
\end{acknowledgements}


\end{document}